\documentclass[runningheads]{llncs}
\usepackage[T1]{fontenc}
\usepackage{graphicx}
\usepackage[utf8]{inputenc} 
\usepackage[T1]{fontenc}    
\usepackage{xcolor}         
\usepackage{hyperref} 
\hypersetup{
    colorlinks=true,       
    linkcolor=blue,       
    citecolor=black,       
    filecolor=black,       
    urlcolor=blue,        
    linkbordercolor=white, 
    citebordercolor=white, 
    filebordercolor=white, 
    urlbordercolor=white   
}
\usepackage{url}            
\usepackage{booktabs}       
\usepackage{amsfonts}       
\usepackage{nicefrac}       
\usepackage{microtype}      
\usepackage{xcolor}         
\usepackage[style=numeric, sorting=none]{biblatex}
\usepackage{tikz}
\usetikzlibrary{shapes,backgrounds}
\usepackage{enumitem}
\usepackage{amsmath,amssymb,amsfonts}
\usepackage{xfrac}
\usepackage{algorithmic}
\usepackage{graphicx}
\usepackage{textcomp}
\usepackage{xcolor}
\usepackage{color,soul}
\usepackage{scalerel}
\usepackage{csquotes}
\usepackage{subcaption}
\usepackage{multirow}
\usepackage{tikz}
\usetikzlibrary{svg.path}
\usepackage[ruled,vlined]{algorithm2e}
\usepackage{todonotes}
\usepackage{graphicx}
\usepackage{nomencl}
\usepackage{comment}

\title{Effects of Dataset Sampling Rate for Noise Cancellation through Deep Learning}

\begin{document}

\author{Brandon Colelough\inst{1}\orcidID{0000-0001-8389-3403} \and
Andrew Zheng\inst{2}}
\authorrunning{B. Colelough et al.}
%

\institute{Department of Computer Science, University of Maryland, 8125 Paint Branch Dr, College Park, MD 20742 
\email{\{brandcol, azheng15\}@umd.edu}}

\maketitle              

\begin{abstract}

\textbf{Background:}
Active noise cancellation has been a subject of research for decades. Traditional techniques, like the Fast Fourier Transform, have limitations in certain scenarios. This research explores the use of deep neural networks (DNNs) as a superior alternative.
\\
\textbf{Objective:}  
The study aims to determine the effect sampling rate within training data has on lightweight, efficient DNNs that operate within the processing constraints of mobile devices.
\\
\textbf{Methods:}  
We chose the ConvTasNET network for its proven efficiency in speech separation and enhancement. ConvTasNET was trained on datasets such as WHAM!, LibriMix, and the MS-2023 DNS Challenge. The datasets were sampled at rates of 8kHz, 16kHz, and 48kHz to analyze the effect of sampling rate on noise cancellation efficiency and effectiveness. The model was tested on a core-i7 Intel processor from 2023, assessing the network's ability to produce clear audio while filtering out background noise.
\\
\textbf{Results:}  
Models trained at higher sampling rates (48kHz) provided much better evaluation metrics against Total Harmonic Distortion (THD) and Quality Prediction For Generative Neural Speech Codecs (WARP-Q) values, indicating improved audio quality. However, a trade-off was noted with the processing time being longer for higher sampling rates. 
\\
\textbf{Conclusions:}  
The Conv-TasNET network, trained on datasets sampled at higher rates like 48kHz, offers a robust solution for mobile devices in achieving noise cancellation through speech separation and enhancement. Future work involves optimizing the model's efficiency further and testing on mobile devices.
\end{abstract}

\section{Introduction} \label{sec:intro}
The advent of mobile communication has revolutionized how we interact, but it has also introduced new challenges, particularly in managing noise in various environments. The surge in mobile device usage in noise-pervasive settings like busy streets or cafes necessitates advanced audio processing solutions to ensure clear and effective communication. We conduct a comparative study on the effects of sampling rates within audio datasets training data for the effectiveness and efficiency of a proven DNN architecture compatible with edge devices. This effort aims to bring robust noise reduction capabilities to common communication tools, thereby improving user experience in noisy environments without excessive computational demands. The field of deep learning-based active noise control (ANC) has seen significant progress, particularly in audio processing for communication systems, biomedical applications, and industrial environments. Key technologies like \href{https://www.nvidia.com/en-us/geforce/guides/nvidia-rtx-voice-setup-guide/}{NVIDIA RTX Voice}, \href{https://krisp.ai/}{Krisp AI} have set the industry standard for noise cancellation, although their specific methodologies remain proprietary. Other computer-based applications such  \href{https://www.utterly.app/}{utterly},  \href{https://product.supertone.ai/clear}{clear by Supertone} and \href{https://magicmic.ai/}{magicmic} and various other computer-based applications are all also available for the problem we set out to solve but, all of these applications require a substantial amount of computational cost and as such are only supported on desktops and are not designed for use on edge devices such as mobile phones. Similar de-noise applications have been developed for use on mobile technology, such as \href{https://apps.apple.com/us/app/denoise-audio-noise-removal/id946423200}{Denoise}, \href{https://apps.apple.com/us/app/audio-noise-reducer-recorder/id1451686645}{Audio Noise Reducer}, \href{https://apps.apple.com/us/app/byenoise-video-audio-editor/id1560151837}{ByeNoise} and many others available right now on the IOS and Android app store. However, Due to the complexity of the de-noise algorithms used by these applications, none of the de-noise applications presently available are able to conduct real-time noise cancellation. Hence, we aim to fill this gap in the present literature and explore systems capable of producing de-noised audio in real-time at high quality on edge devices such as a mobile phone.  The development of ANC technologies, particularly through algorithms like Filtered-x Least Mean Square (FxLMS), has significantly enhanced electroacoustic systems by addressing challenges such as acoustic feedback and secondary route estimation \cite{Lu2021}. These advancements extend to Nonlinear ANC (NLANC), employing methods based on the Volterra series, Hammerstein models, and Functional Link Artificial Neural Networks (FLANN), which show promise in managing the complexities inherent to nonlinear environments \cite{Lu2021a}. Additionally, heuristic algorithms like genetic and particle swarm optimizations have been introduced to tackle these nonlinear challenges \cite{Lu2021a}.In practical applications, Microsoft Teams incorporates AI-driven ANC to improve audio clarity in video calls, likely utilizing neural networks like CNNs and RNNs to filter speech from background noise \cite{Sharma2023}. In the biomedical field, innovative real-time noise cancellation techniques use deep learning to minimize EMG interference in EEG signals, suggesting applications in both consumer electronics and industrial monitoring \cite{Porr2020}. Advances in vocoder technology through deep learning, such as the integration of WaveNet-inspired CNNs and DSP techniques, enhance audio quality and enable effective speech synthesis, showing significant improvements in systems like the NVSR model and the "Vogen Voc" \cite{Chandna2019, Liu2022, Yang2023}. VocBench and LightVoc represent the latest in vocoder testing and neural coding, respectively, highlighting rapid progress in speech synthesis and audio processing \cite{AlBadawy2021, Dang2023}. Deep learning continues to outperform traditional methods in ANC across various environments, from personal devices to challenging industrial settings, proving its efficacy and adaptability \cite{AkarshS.2022, Kejalakshmi2022, Zhang2020, Zhang2021, Mostafavi2023, Suresh2020}.

\section{Related Work}\label{related_work:noise_sep}

\subsection{Audio Source Separation}\label{subsec:noise_sep}
Deep learning techniques have been a major driving force behind recent advances in audio source separation, which have resulted in notable developments in this field.  \cite{Stoller2018} presented the Wave-U-Net architecture, a novel method for separating audio sources that overcome the constraints associated with phase information and fixed spectrum transformations present in conventional spectrogram-based techniques by converting the U-Net model to a one-dimensional format and concentrating on time-domain processing. The  Wave-U-Net model offers a potential direction for high-quality separation in a variety of audio settings by handling long-range temporal dependencies efficiently. \cite{Takahashi2017} presented a Multi-Scale Multi-Band DenseNet (MMDenseNet) to address large input and output dimensions and extend the DenseNet architecture to support lengthy context modelling. The success of MMDenseNet in the Signal Separation Evaluation Campaign (SiSEC) 2016 is especially indicative of the improved separation performance made possible by their unique multi-band methodology, in which each frequency band is simulated independently. The SUccessive DOwnsampling and Resampling of Multi-Resolution Features (SuDoRM-RF) network is a novel design for universal audio source separation, created by \cite{Tzinis2020}. SuDoRM-efficient RF's architecture considerably reduces processing demands by combining one-dimensional convolutions with downsampling and resampling.

\subsection{Speech Separation}\label{subsec:Speech_separation}
Speech Separation is a subset of audio source separation that can be used for the implementation of a technique to eliminate background noise from natural speech. \cite{Lu2021a} most recently presented a review work that offers a thorough summary of how deep learning might improve supervised speech separation approaches and procedures. Their study describes the development and innovations in deep learning for speech separation, emphasizing improvements in acoustic modelling approaches and algorithmic tactics through 2018. However, after the publication of this overview report in 2018, significant advancements have been made in the literature. \cite{Subakan2020} proposed the SepFormer architecture, which provides state-of-the-art speech separation performance, but, its size prevents it from being deployed on an edge device. Similarly \cite{DellaLibera2022} describe a RE-SepFormer design that reduces computational costs but maintains substantially but not enough for implementation on edge devices. The MossFormer2 architecture presented by \cite{Zhao2023} shows promise for network size and cost with its hybrid model, but,  the implementation details on their architecture lack explicit details on the computational requirements during the forward pass, which is crucial for edge device implementation. For deep neural network archetypes used in noise cancellation, the designs presented by \cite{Zhang2021} and \cite{Porr2020} offer the best performance. Nevertheless, in real-world scenarios, these implementations are too slow for a less capable edge device to use. Two research studies that provide models that meet our time limitations for edge device functioning were found through a more thorough analysis of pertinent literature. We found that both the convolutional architecture explored in the Conv-TasNET architecture, detailed by\cite{Luo2019a}, and the Skip Memory LSTM model presented by \cite{Li2022} with SKIM performed well. However, Conv-TasNET was superior in terms of effectiveness for speech separation and audio enhancement and demonstrated greater efficiency in the forward passage of the network.

\subsection{Speech Enhancement}\label{subsec:noise_enh}
Deep learning approaches have once again changed the game in the speech enhancement space, producing breakthroughs in speech separation (AV-SS) and audio-visual speech enhancement (AV-SE). A detailed review is given by \cite{Michelsanti2021}, which emphasizes the combination of audio signals and visual clues such as lip movements. This multi-modal strategy has been successful in improving voice quality by utilizing simple neural network designs such as feedforward and recurrent neural networks. Specific training targets and objective functions, fusion procedures, and auditory and visual elements are essential components of these systems.  \cite{Chhetri2023} offers another thorough analysis that divides different speech improvement methods into four categories: time domain, statistical-based, transform domain, and AI-based. The survey by  \cite{Chhetri2023} addresses the need for effective algorithms required for applications such as noise reduction and investigates several approaches such as Wiener filtering, comb filtering, and deep learning techniques. This survey highlights how speech augmentation is widely used in a variety of industries, including automotive, medical, and telecommunication, demonstrating the field's increasing importance. Although not directly related to speech enhancement, the study "Proximal Policy Optimization Algorithms" by \cite{Schulman2017} offers insightful information about effective algorithmic frameworks surrounding the speech separation field and specifically presents PPO, a policy gradient approach to reinforcement learning that emphasizes the significance of optimizing objective functions and stability in performance for voice enhancement and provides a more straightforward and effective approach than earlier techniques used within the field. \cite{Pascual2017} introduce SEGAN, the  Speech Enhancement Generative Adversarial Network,  a generative adversarial framework operating on raw audio waveforms. This end-to-end model effectively handles multiple noise types and speaker variations, illustrating the potential of GANs in speech processing. The \enquote{Phase-Aware Speech Enhancement with Deep Complex U-Net}  architecture presented by \cite{Choi2019} addresses the difficulty of phase estimation in speech enhancement through their introduction of Deep Complex U-Net (DCUnet). DCUnet addressed a crucial issue that is sometimes overlooked in traditional models by combining complex-valued operations with an innovative loss function to maximize efficiency. \cite{Hu2020} presents the \enquote{Deep Complex Convolution Recurrent Network for Phase-Aware Speech Enhancement}  (DCCRN) for specific use in the context of real-time speech augmentation. The DCCRN model demonstrated the efficiency and efficacy of phase-aware voice augmentation in real-time processing scenarios by integrating recurrent structures and complex-valued convolution.  A convolutional neural network design is proposed by \cite{Park2016}, with an emphasis on babbling noise that is frequently present in hearing aids. \cite{Park2016} demonstrates how CNNs outperform conventional neural network models in speech enhancement tests due to their efficiency in parameter usage. The creation of FullSubNet, which combines full-band and sub-band models for real-time single-channel speech enhancement, by \cite{Hao2020} has made a substantial contribution to the area. Their novel approach demonstrates how full-band and sub-band information complements each other to improve voice quality by efficiently handling a variety of loud environments and reverberation effects. Lastly, the ConvTasNet architecture from \cite{Luo2019a} also introduced speech enhancement capabilities that operate directly in the time domain. The ConvTasNet architecture is a lightweight and fast network that outperforms conventional time-frequency magnitude masking techniques and is perfect for real-time applications due to its efficient architecture and usage of Temporal Convolutional Networks (TCNs), which enable accurate speech separation with smaller model sizes and lower latency.

\subsection{Available Datasets}\label{subsec:datasetl}
The ongoing research within the domain of audio separation, speech separation and speech enhancement is significantly influenced by the availability of diverse datasets. \cite{Hershey2015a} developed a foundational dataset based on the Wall Street Journal corpus (WSJ0), applying a 'Deep Clustering' method for acoustic source separation that maps spectrogram features into an embedding space. The Deep Clustering method shows that the model can generalize beyond its training data, and is especially good at managing combinations of multiple speakers. By adding ambient noise recordings from actual situations to the wsj0-2mix dataset, the WHAM! dataset by \cite{Wichern2019} improves realism and tests speech separation algorithms in more challenging, noisy settings. Adding to this, the WHAMR! dataset from \cite{Maciejewski2019} presents a more challenging scenario of speech recorded in noisy and reverberant environments by introducing synthetic reverberation. The AISHELL-4 dataset by \cite{Fu2021} offers extensive Mandarin speech data, recorded in conference settings with an 8-channel circular microphone array. Based on LibriSpeech, LibriMix by \cite{Cosentino2020} includes speech data mixtures from one, two, and three speakers that were recorded at 8 kHz and 16 kHz. It emphasizes perceptually balanced mixtures and natural conversation dynamics, and allows users to mix in WHAM! noise. Scalable noisy speech data is made available by Microsoft's MS-SNSD dataset \cite{Reddy2019a}, which makes training and testing deep learning models for speech augmentation easier. The ICASSP 2023 Deep Noise Suppression Challenge \cite{Dubey2023} featured the most recent version of a Microsoft dataset for speech separation and enhancement, highlighting the field's progress by emphasizing deep speech enhancement models for denoising, dereverberation, and interference suppression in complex audio environments.

\section{Methodology}\label{sec:methodology}

\subsection{Architecture Design}\label{subsec:arch_design}
The Conv-TasNET framework \cite{Luo2019a} was utilised to determine the effect of the dataset sampling rate on model performance. This choice is driven by Conv-TasNET’s proven efficiency in speech separation and speech enhancement, making it ideal for processing within the limited computational capacities typical of mobile devices. In addition, a CNN-driven model is optimal for data that forward progressed in a single parse as a large matrix, where the location of the data to other nearby data is meaningful. Conv-TasNET is also suitable for us because of the minimal complexity of the model, allowing for optimal use on mobile devices. 
 
\begin{table}[ht]
\centering
\begin{tabular}{lccc}
\hline
\multicolumn{4}{c}{\textbf{Hyperparameters used for training the network}} \\
\hline
& \textbf{8kHz} & \textbf{16kHz} & \textbf{48kHz} \\
\hline
\textbf{Data} & & & \\
\hline
num sources & 1 & 1 & 1 \\
sample rate & 8000 & 16000 & 48000 \\
segment & 3 & 3 & 3 \\
\hline
\textbf{Training} & & & \\
\hline
batch size & 24 & 12 & 1 \\
epochs & 100 & 50 & 50 \\
num workers & 4 & 4 & 4 \\
\hline
\end{tabular}
\caption{Hyperparameters used for training the network across different sample rates. Note that parameters that remained constant across each model are not shown.}
\label{table:proj_arch_params}
\end{table}

\subsection{Dataset and hyper-parameters}\label{subsec:dataset}
The network was trained using three distinct datasets, sampled at 8kHz, 16kHz and 48kHz, to determine the effect that the sampling rate has on the produced audio quality. The network was trained through a high-performance computing cluster utilising a parallel computing environment that contained 4x RTX-A6000 GPUs with 16 CPUs and 128GB of RAM. The model was adapted to enhance single-channel audio and produce only speech audio from noisy audio. The network hyper-parameters used for training are shown in table \ref{table:proj_arch_params}. As the network size and complexity increased substantially with an increased sampling rate, the batch size and number of epochs trained varied across the three sampling rates utilised due to computational limits. All other network training variables were standardized across all experimental trials to ensure uniformity in testing conditions and maintain the integrity of the comparative analysis. 

\subsubsection{Librimix with WHAM! (8kHz and 16kHz sampling rate)}\label{subsubsec:dataset-LibWham}
To establish a baseline, we trained the architecture on two datasets made from LibriMix speech audio library \cite{Cosentino2020} with noise mixed in from the  WHAM! dataset \cite{Wichern2019}. The mixture generated from Librimix utilised focused on an 8kHz or 16kHz sampling rate blend with 2-speaker audio for speech separation. A 360-hour dataset of noisy speech audio for the 8kHz sampling rate and a 100-hour dataset of noisy speech audio for the 16kHz sampling rate were hence produced and utilised for the training of the model to gather a baseline. A training, evaluation, and testing split of 80:10:10 was utilised for both the 8kHz and 16kHz sampling rates.  

\subsubsection{MS-DNS Challenge (48kHz sampling rate)}\label{subsubsec:dataset-MS-DNS}
The model was further trained using a novel 48kHz sampling rate mixture from the ICASSP 2023 challenge dataset presented by \cite{Dubey2023}. A 150-hour dataset of noisy speech audio for the 48kHz sampling rate was hence produced and an 80:10:10 training, evaluation, and testing split was utilised for model training and evaluation. The Track 2-Speakerphone dataset from the challenge was selected for the model training. 

\subsection{Metrics for Evaluation}\label{method:model_effectiveness}
For model efficiency, we analyse the time and computational complexity of the forward progression of the model. Model effectiveness was measured against four metrics including the Scale-Invariant Signal Distortion Ratio (SI-SDR),  Short-Time Objective Intelligibility (STOI), Total Harmonic Distortion (THD) and Quality Prediction For Generative Neural Speech Codecs (WARP-Q). The SI-SDR and STOI are standard metrics used for the field of speech separation and are calculated in alignment with the methods produced by the asteroid team \cite{Pariente2020Asteroid}. THD is a standard method for measuring audio quality within the domain of signal processing and methods for calculation align with methods shown by \cite{DSP_book}. A hamming window was first used to extract the frequency domain signal from the audio-separated signal. Fundamental and harmonic frequencies were isolated, and THD was calculated as the ratio of the root mean square of the harmonic frequencies to the fundamental frequency, presented as a percentage. Lastly, the Warp-Q metric computes the distance between reference and degraded signals using dynamic time warping on spectral features as explained by \cite{Wissam_IET_Signal_Process2022}. Both the THD and WARP-Q were then normalised against the original, noiseless audio sampled at 48kHz. The THD normalisation score was taken as the distortion present before processing minus the distortion present after processing:

\[
\text{THD}_{\text{norm}} = \text{signal}_{\text{computed}} - \text{signal}_{\text{ref}}
\]

The normalization score was calculated as the inverse of the absolute difference plus one, where closeness to the reference signal yields a higher score:

\[
\text{WARP-Q}_{\text{norm}} = \frac{1}{|\text{signal}_{\text{computed}} - \text{signal}_{\text{ref}}| + 1}
\]

\section{Results}\label{sec:results}

\subsection{Model Demonstration}\label{sec:efficiency}
The noisy audio clip is sampled at 48kHz and was hence downsampled to be run through the 8kHz and 16kHz models before being again upsampled to 48kHz post-processing. Due to the downsampling and upsampling that was conducted with the 8khz and 16kHz models, these two models produce an overall grainier audio quality.\footnote{Audio samples produced by the model can be found on the project's GitHub page: \href{https://github.com/Brandonio-c/ClearComm-NN}{ClearComm-NN}}. 

\begin{table}[ht]
\centering
\resizebox{0.5\textwidth}{!}{%
\begin{tabular}{lccc}
\hline
\multicolumn{4}{c}{\textbf{Model Effectiveness Results}} \\
\hline
& \textbf{8kHz} & \textbf{16kHz} & \textbf{48kHz} \\
\hline
\textbf{SI-SDR} & 14.70dB & 14.74 dB & 14.92 dB \\
\textbf{STOI} & 92.60\% & 93.11\% & 86.36\% \\
\textbf{THD} & 41.09\% & 24.59\% & 2.21\% \\
\textbf{WARP-Q} & 38.40\% & 58.38\% & 77.94\% \\
\hline
\end{tabular}
}
\caption{Effectiveness results of the ConvTasNet model trained on similar datasets across different sample rates of 8kHz, 16kHz and 48kHz. The forward progression of the model was conducted on a single Intel core-i7 CPU}
\label{table:model_effectiveness_results}
\end{table}

\begin{figure}[ht]
    \centering
    \begin{minipage}{0.48\textwidth}
        \centering
        \includegraphics[width=\linewidth]{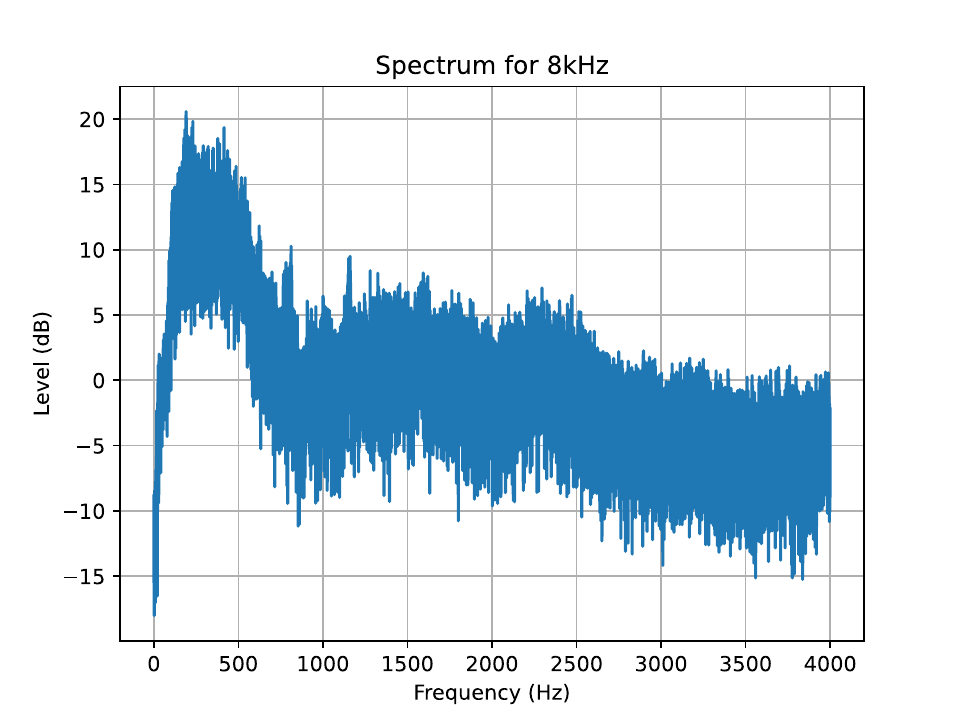} 
        \caption*{8 kHz Sampling rate}
    \end{minipage}\hfill
    \begin{minipage}{0.48\textwidth}
        \centering
        \includegraphics[width=\linewidth]{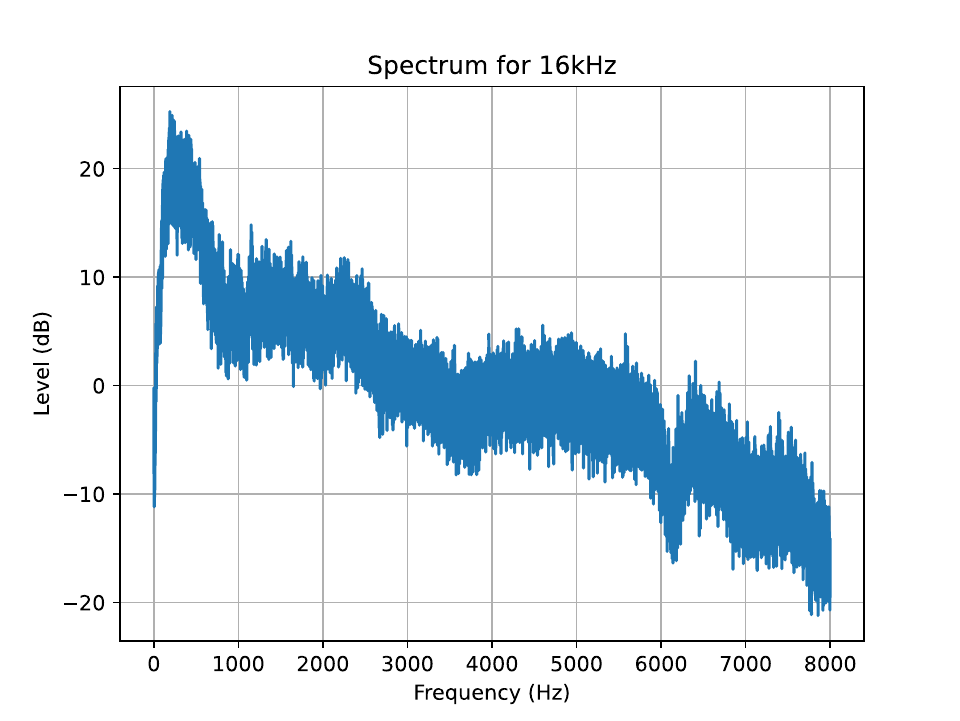} 
        \caption*{16 kHz Sampling rate}
    \end{minipage}
    \begin{minipage}{0.48\textwidth}
        \centering
        \includegraphics[width=\linewidth]{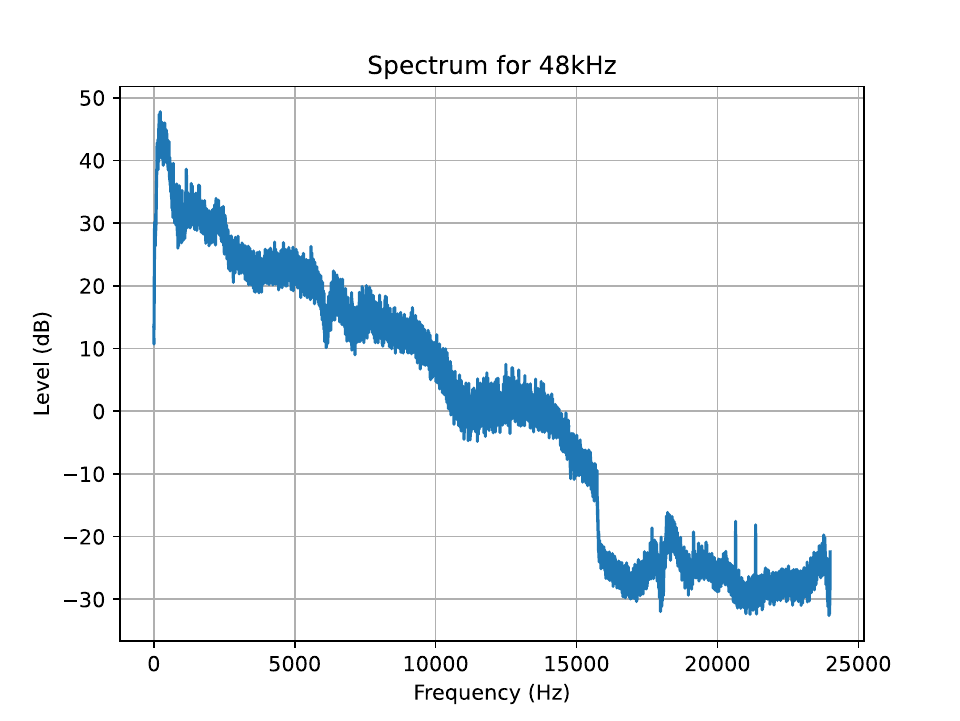} 
        \caption*{48 kHz Sampling rate}
    \end{minipage}\hfill
    \begin{minipage}{0.48\textwidth}
        \centering
        \includegraphics[width=\linewidth]{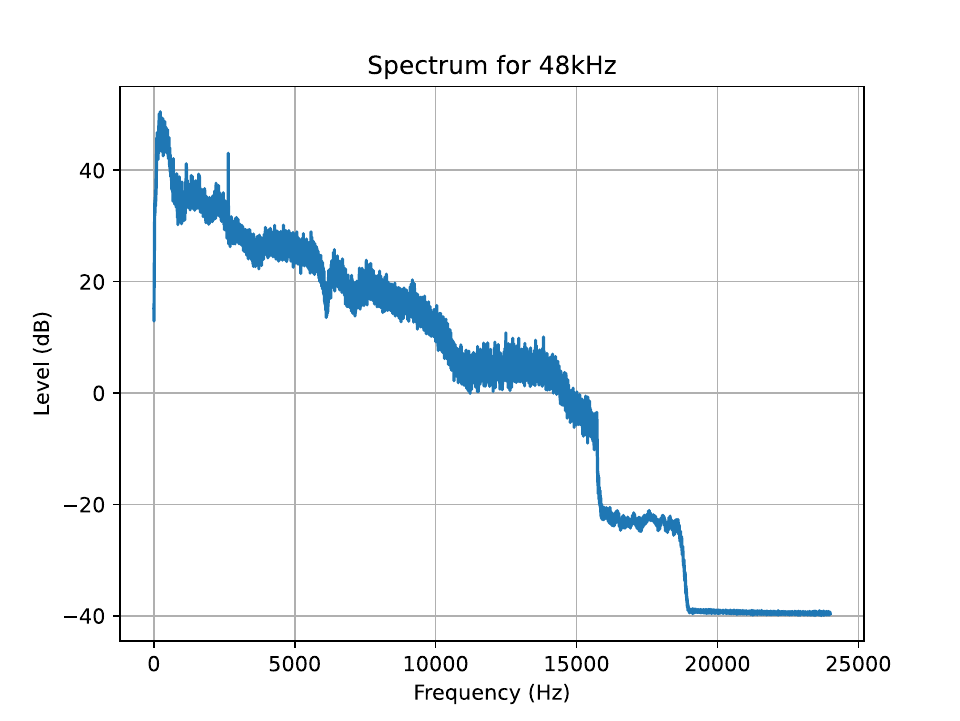} 
        \caption*{48 kHz Reference signal spectrum}
    \end{minipage}
    \caption{The noisy one-minute audio clip shown above was taken and passed through the models trained on 8kHz, 16kHz and 48kHz. The frequency spectrum of the noise-filtered audio files is shown above as well as the original, noiseless audio files.}
    \label{fig:spectrum}
\end{figure}

\begin{figure}[ht]
    \centering
    \begin{minipage}{0.48\textwidth}
        \centering
        \includegraphics[width=\linewidth]{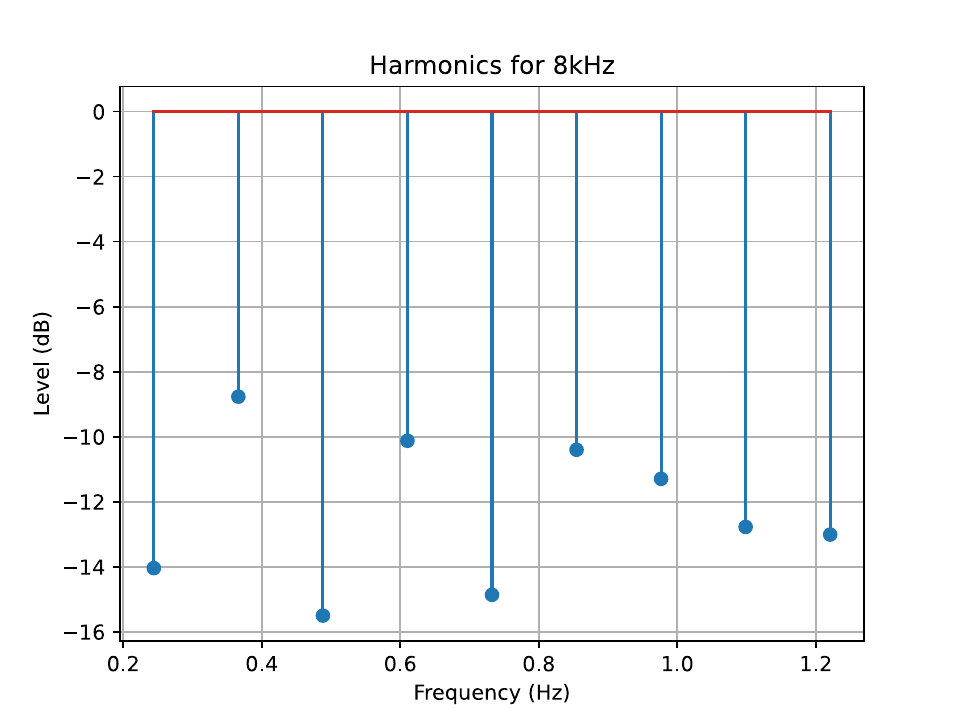} 
        \caption*{8 kHz Harmonics}
    \end{minipage}\hfill
    \begin{minipage}{0.48\textwidth}
        \centering
        \includegraphics[width=\linewidth]{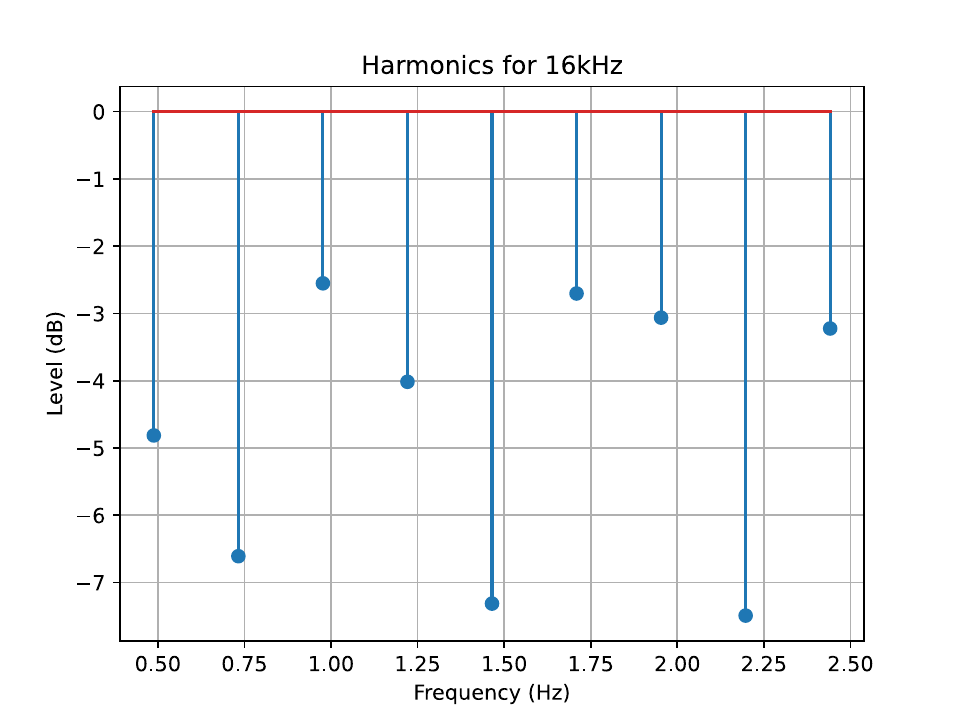} 
        \caption*{16 kHz Harmonics}
    \end{minipage}
    \begin{minipage}{0.48\textwidth}
        \centering
        \includegraphics[width=\linewidth]{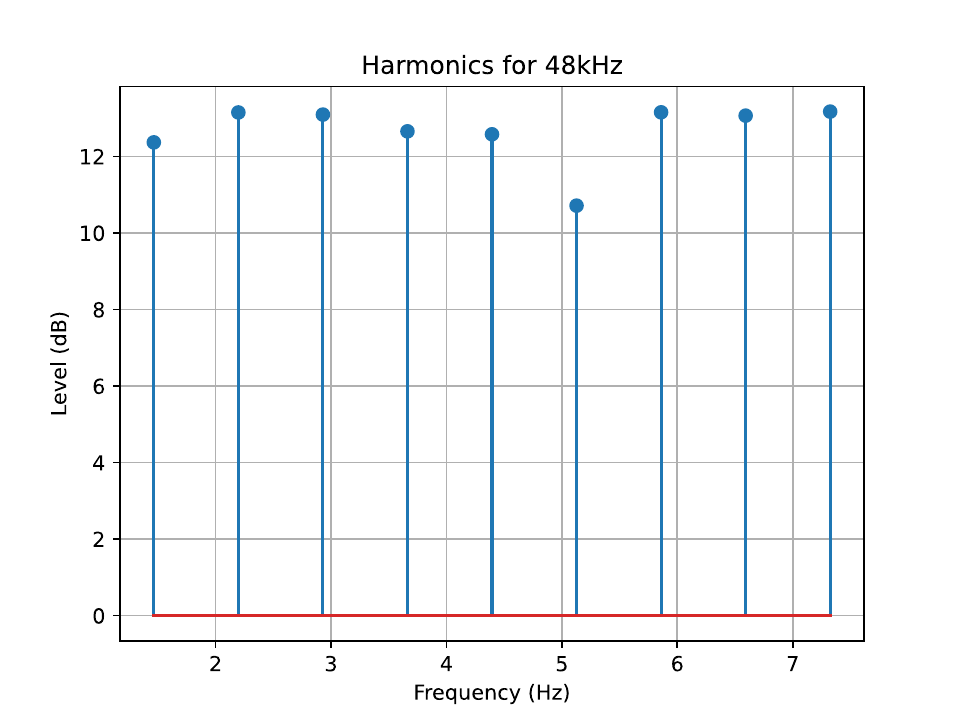} 
        \caption*{48 kHz Harmonics}
    \end{minipage}\hfill
    \begin{minipage}{0.48\textwidth}
        \centering
        \includegraphics[width=\linewidth]{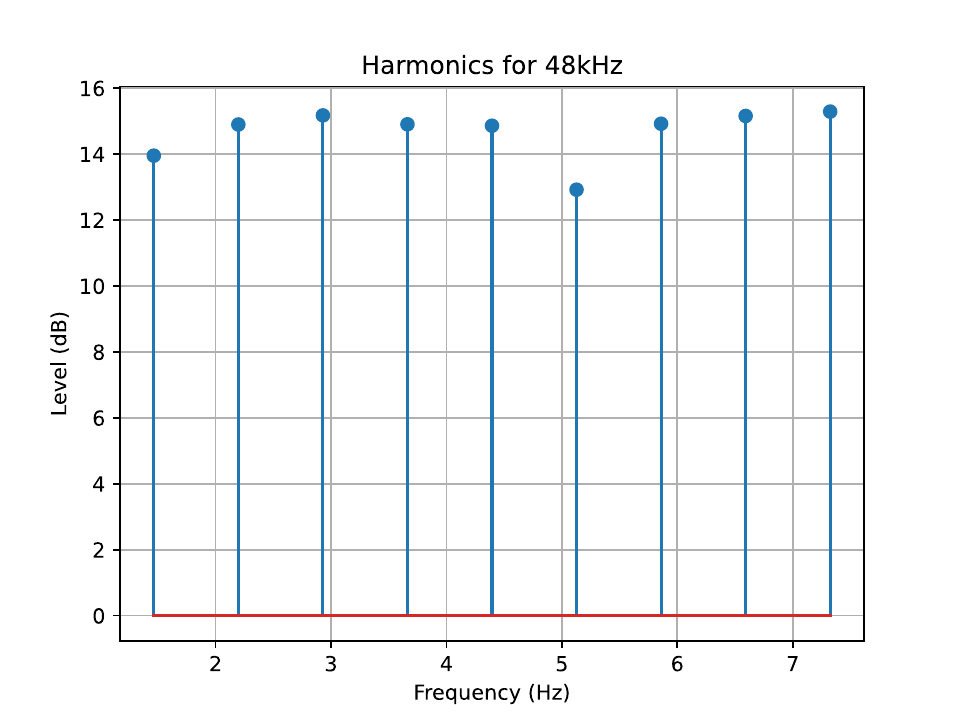} 
        \caption*{48 kHz Original Harmonics}
    \end{minipage}
    \caption{Harmonic content comparison across different sampling rates and the original signal for audio quality assessment}
    \label{fig:harmonics}
\end{figure}

\subsection{Effectiveness of model:}\label{sec:effective}
Results for SI-SDR, STOI, THD and WARP-Q are all shown in table \ref{table:model_effectiveness_results}. Figure \ref{fig:spectrum} and Figure \ref{fig:harmonics} show the frequency spectrum and harmonics of the de-noised audio clips produced by the 8kHz, 16kHz, 48kHz and the original speech reference signal. The 8khz and 16kHz were upsampled to a sampling rate of 48kHz utilising a Polynomial Interpolation upsampling function before analysis.

\subsection{Efficiency of model:}\label{sec:efficiency}
\begin{table}[ht]
\centering
\resizebox{0.5\textwidth}{!}{%
\begin{tabular}{lccc}
\hline
\multicolumn{4}{c}{\textbf{Model Efficiency Results}} \\
\hline
& \textbf{8kHz} & \textbf{16kHz} & \textbf{48kHz} \\
\hline
\textbf{10s clip} &  118 ms & 238 ms & 724ms \\
\textbf{5s clip} & 63 ms & 121 ms & 754 ms \\
\textbf{1s clip} & 18 ms & 34 ms & 102 ms \\
\hline
\end{tabular}
}
\caption{Efficiency results of the ConvTasNet model trained on similar datasets across different sample rates of 8kHz, 16kHz and 48kHz. The forward progression of the model was conducted on a single Intel core-i7 CPU}
\label{table:model_efficiency_results}
\end{table}

Table \ref{table:model_efficiency_results} shows the forward processing of the model to produce a de-noised clip from noisy audio. The table shows forward inferencing for the 8kHz, 16kHz and 48kHz models conducted on a single Intel core i7 CPU for 1s, 5s and 10s noisy audio clips. 

\section{Discussion}\label{sec:discussion}

\subsection{Effectiveness of model}\label{subsec:disc_effectiveness}
The SI-SDR assesses the ratio between the signal (in our case the separated speech) to distortion (in our case added noise) while being scale-invariant, thus independent of signal loudness or scale and the STOI predicts speech intelligibility, evaluating how understandable the speech is to a typical listener making the SI-SDR and STOI good metrics to determine whether the model can effectively produce the desired speech audio whilst filtering out the background noise. The THD measures the degree to which a system distorts the harmonic components of the signal compared to its fundamental frequency, and the WARP-Q assesses the perceptual quality of the generated speech, taking into account factors like naturalness and clarity. Together, these metrics ensure that not only is the speech intelligible and clear of noise (as measured by SI-SDR and STOI), but it also retains fidelity to the original sound without introducing unintended harmonic distortions ( as measured by the THD and WARP-Q). The SI-SDR values, as shown in Table \ref{table:model_effectiveness_results}, indicate a consistent performance across 8 kHz and 16 kHz sample rates, with a slight increase at 48 kHz (14.70 dB, 14.74 dB, and 14.92 dB, respectively). Whilst there as not a notable increase in SI-SDR, it should also be noted here that the higher sampling rate datasets were trained over half of the epochs due to computational complexity. In contrast, the STOI results, which range from 92.60\% at 8 kHz to 86.36\% at 48 kHz, imply a slight decline in how well typical listeners might understand the speech as the sample rate increases. The decline in performance for the 48kHz model for STOI is likely due to the lowered amount of training conducted (i.e. lower epoch size) and the reduction in batch size due to computational constraints during model training. However, there is a drastic improvement in THD from 41.09\% at 8 kHz to 2.21\% at 48 kHz suggesting that the model increasingly preserves the harmonic integrity of the audio signal as the sample rate increases. Lower THD at higher sample rates indicates less harmonic distortion, which is critical for audio quality in high-fidelity applications. Similarly, the WARP-Q results also show a progressive improvement (38.40\% at 8 kHz to 77.94\% at 48 kHz), underscoring a better perceptual audio quality at higher sample rates. This suggests that the model not only improves in handling the nuances of audio signals but also enhances the naturalness and clarity, making the output more pleasing and accurate to the human ear. The frequency spectrum and harmonic content further substantiate these findings. As illustrated in Figure \ref{fig:spectrum}, the audio clip denoised at 48 kHz more closely mirrors the reference signal's frequency spectrum, indicating superior noise filtering capabilities. This alignment suggests that the model, when operated at 48 kHz, more effectively suppresses noise while preserving essential speech characteristics. Similarly, Figure \ref{fig:harmonics} reveals that the fundamental frequencies and harmonics of the 48 kHz output are significantly closer to the reference, an average 1dB difference, compared to the outputs at 8 kHz and 16 kHz which ranged from a 20dB to 30dB difference from the reference speech signal. This closer resemblance in harmonic structure from the audio produced by the model trained on higher sample rates corroborates the model’s capacity to maintain audio integrity, echoing the findings from THD and WARP-Q metrics.

\subsection{Efficiency of model}\label{subsec:disc_effectiveness}

\cite{Luo2019a} report a forward inference time of 0.4ms for a single frame when the ConvTasNet architecture is deployed on one core-i7 Intel CPU, where a frame is given as the frame size divided by the sampling rate, i.e. for a sample chunking size of 256 samples per frame, one frame sampled at 8kHz would equate to 32ms of audio and is processed at 0.4ms. The forward processing time is hence given by:
\begin{equation}
    \text{Processing time} = \frac{N}{n} \times 0.4 \, \text{ms}
\end{equation}

Where N is the number of samples per frame and n is the sampling rate selected. Theoretically, using 256 samples per frame, this then equates to roughly 12.5ms to process 1s of audio at 8kHz, 25ms at 16kHz and 75ms at 48kHz. As shown in table \ref{table:model_efficiency_results}, per every second of audio chunk processed, the 8kHz model has a forward progression time of roughly 18ms, the 16kHz model 34ms and the 48kHz model 102ms. The minimal difference in measure and theoretical results is likely due to environmental factors. As observed, doubling the model size from 8kHz to 16 kHz adds only 16ms forward processing time due to manageable data growth, but the sixfold increase to 48 kHz from 8kHz exponentially intensifies computation and memory usage, causing a 5x slowdown, as the model's architecture and complexity struggle to handle such data volume. The \enquote{average audio leading threshold for a/v sync detection is 185.19 ms} \cite{Younkin08} and as such, the present model is likely efficient enough to function in real-time running on a mobile device on a single CPU given the mobile phone processor is at least a quad-core. Alternatively, to speed up the forward processing time, processing across multiple CPUs or dedicated hardware such as a GPU/TPU could be achieved on some mobile devices. For less computationally available hardware such as older devices, a reduction to the layer size of the present model could also be implemented. Additionally, training the model on exactly eh Nyquist rate would also increase efficiency while maintaining effectiveness. However, Further testing should be conducted on a range of mobile devices to test the above assumptions.

\subsection{Model trade-off}
There was an observed increase in the computation time required to separate audio with the model trained utilising a 48kHz sampling rate. However, there was also an observed reduction in audio quality produced by the models trained at 8kHz and 16kHz attributed to the necessary processes of downsampling and upsampling involved with lower sampling rates. Downsampling to 8kHz inadequately captures the range of frequencies discernible by the human ear as human hearing typically ranges from 20Hz to 20kHz, and to accurately reproduce this range, a sampling rate of at least twice the maximum frequency (the Nyquist rate) is required. As such, the industry standard for high-quality audio is set at 44.1kHz, which is sufficient to cover the audible spectrum with minimal loss of fidelity. In addition, models trained on 16kHz data greatly hinder the model's effectiveness on standard 44.1kHz data that we want our model to denoise since the implicit assumption of the time between each sample in the signal is changed by passing a 44.1kHz signal through the model. The time between each sample in the original sample in a 44.1 kHz signal is different from the 8kHz data that the model is trained on. Hence, training our network to accommodate audio sources at 44.1kHz would effectively mitigate the need for downsampling and subsequent upsampling, thus preserving the integrity of the audio signal. This approach aligns with industry standards and ensures that the processed audio remains within the optimal range for human hearing. By implementing training at this higher sampling rate, we show a significant improvement in the quality of the separated audio, and align our model’s performance more closely with real-world applications and user expectations.

\section{Conclusion}\label{sec:conclusion}
We aimed to determine the effect that dataset sampling rate had on a lightweight, highly effective Deep Neural Network that could conduct real-time noise cancellation for mobile voice communications in noisy environments. To achieve this, we conducted a comparative study on the effect that the audio sampling rate had on an existing model architecture for effectiveness and efficiency. Our review of the literature indicated a significant body of work in the field of deep learning-based noise cancellation, with specific emphasis on innovations that have transformed audio processing for a variety of applications in the deep learning domain, especially in the area of speech separation. We chose the Conv-TasNET network architecture as our foundation model to accomplish noise cancellation on an edge device. To train the Conv-TasNet architecture, we chose the WHAM!, LibriMix, and MS-2023 DNS challenge datasets for their accessibility and dataset features which aligned with our goals. The model was hence trained on a mix of these datasets utilising a sampling rate of 8kHz, 16kHz and 48kHz respectively. Our findings indicate that increased model effectiveness to produce high-quality separated audio from the noisy speech was achieved through training on data with a higher sampling rate via an algorithm for speech enhancement. In addition, the model showed robustness towards audio signals measured naturally despite being trained on only generated audio signals. The 48kHz model is likely efficient enough to achieve noise cancellation through audio source separation and enhancement on mobile phones, however, implementation of such a noise-enhancing feature in real-time on edge devices needs to be created for accurate testing.

\printbibliography

@InProceedings{Suresh2020,
  author = {Ananda Theertha Suresh and Asif Khan},
  title  = {A COMPARISON OF NOISE CANCELLATION OF SPEECH SIGNALS USING ADAPTIVE LMS ALGORITHM AND DEEP LEARNING ALGORITHM},
  year   = {2020},
  url    = {https://api.semanticscholar.org/CorpusID:216653478},
}

@Article{Sharma2023,
  author       = {Sharma, Pawankumar and Dash, Bibhu},
  journal      = {AIRCC’s International Journal of Computer Science and Information Technology},
  title        = {Active Noise Cancellation in Microsoft Teams Using AI amp; NLP Powered Algorithms},
  year         = {2023},
  month        = {Feb.},
  number       = {1},
  pages        = {31–42},
  volume       = {15},
}

@InProceedings{Kejalakshmi2022,
  author = {Dr V. Kejalakshmi and A. Kamatchi and M. A. Anusuya},
  title  = {Active Noise Cancellation using Deep learning},
  year   = {2022},
  url    = {https://api.semanticscholar.org/CorpusID:250616211},
}

@InProceedings{Zhang2020,
  author     = {Zhang, Hao and Wang, DeLiang},
  booktitle  = {Interspeech 2020},
  title      = {A Deep Learning Approach to Active Noise Control},
  year       = {2020},
  month      = oct,
  publisher  = {ISCA},
  series     = {interspeech 2020},
  collection = {interspeech 2020},
  doi        = {10.21437/interspeech.2020-1768},
}

@Article{Yang2023,
  author    = {Yang, Runxuan and Peng, Yuyang and Hu, Xiaolin},
  journal   = {IEEE/ACM Transactions on Audio, Speech, and Language Processing},
  title     = {A Fast High-Fidelity Source-Filter Vocoder With Lightweight Neural Modules},
  year      = {2023},
  issn      = {2329-9304},
  pages     = {3362--3373},
  volume    = {31},
  doi       = {10.1109/taslp.2023.3321191},
  publisher = {Institute of Electrical and Electronics Engineers (IEEE)},
}

@Misc{Fu2021,
  author    = {Fu, Yihui and Cheng, Luyao and Lv, Shubo and Jv, Yukai and Kong, Yuxiang and Chen, Zhuo and Hu, Yanxin and Xie, Lei and Wu, Jian and Bu, Hui and Xu, Xin and Du, Jun and Chen, Jingdong},
  title     = {AISHELL-4: An Open Source Dataset for Speech Enhancement, Separation, Recognition and Speaker Diarization in Conference Scenario},
  year      = {2021},
  copyright = {Creative Commons Attribution 4.0 International},
  doi       = {10.48550/ARXIV.2104.03603},
  publisher = {arXiv},
}

@Article{Lu2021a,
  author    = {Lu, Lu and Yin, Kai-Li and de Lamare, Rodrigo C. and Zheng, Zongsheng and Yu, Yi and Yang, Xiaomin and Chen, Badong},
  title     = {A survey on active noise control in the past decade–Part II: Nonlinear systems},
  journal   = {Signal Processing},
  year      = {2021},
  issn      = {0165-1684},
  month     = apr,
  pages     = {107929},
  volume    = {181},
  doi       = {10.1016/j.sigpro.2020.107929},
  publisher = {Elsevier BV}
}

@Misc{Lu2021,
  author    = {Lu, Lu and Yin, Kai-Li and de Lamare, Rodrigo C. and Zheng, Zongsheng and Yu, Yi and Yang, Xiaomin and Chen, Badong},
  title     = {A survey on active noise control techniques -- Part I: Linear systems},
  year      = {2021},
  copyright = {arXiv.org perpetual, non-exclusive license},
  doi       = {10.48550/ARXIV.2110.00531},
  publisher = {arXiv}
}

@Misc{Subakan2020,
  author    = {Subakan, Cem and Ravanelli, Mirco and Cornell, Samuele and Bronzi, Mirko and Zhong, Jianyuan},
  title     = {Attention is All You Need in Speech Separation},
  year      = {2020},
  copyright = {arXiv.org perpetual, non-exclusive license},
  doi       = {10.48550/ARXIV.2010.13154},
  publisher = {arXiv},
}

@InProceedings{Chandna2019,
  author    = {Chandna, Pritish and Blaauw, Merlijn and Bonada, Jordi and Gomez, Emilia},
  booktitle = {ICASSP 2019 - 2019 IEEE International Conference on Acoustics, Speech and Signal Processing (ICASSP)},
  title     = {A Vocoder Based Method for Singing Voice Extraction},
  year      = {2019},
  month     = may,
  publisher = {IEEE},
  doi       = {10.1109/icassp.2019.8683323},
}

@Article{Luo2019a,
  author    = {Luo, Yi and Mesgarani, Nima},
  journal   = {IEEE/ACM Transactions on Audio, Speech, and Language Processing},
  title     = {Conv-TasNet: Surpassing Ideal Time–Frequency Magnitude Masking for Speech Separation},
  year      = {2019},
  issn      = {2329-9304},
  month     = aug,
  number    = {8},
  pages     = {1256--1266},
  volume    = {27},
  doi       = {10.1109/taslp.2019.2915167},
  publisher = {Institute of Electrical and Electronics Engineers (IEEE)},
}

@Article{Zhang2021,
  author    = {Zhang, Hao and Wang, DeLiang},
  journal   = {Neural Networks},
  title     = {Deep ANC: A deep learning approach to active noise control},
  year      = {2021},
  issn      = {0893-6080},
  month     = sep,
  pages     = {1--10},
  volume    = {141},
  doi       = {10.1016/j.neunet.2021.03.037},
  publisher = {Elsevier BV},
}

@Article{Mostafavi2023,
  author    = {Mostafavi, Alireza and Cha, Young-Jin},
  journal   = {Automation in Construction},
  title     = {Deep learning-based active noise control on construction sites},
  year      = {2023},
  issn      = {0926-5805},
  month     = jul,
  pages     = {104885},
  volume    = {151},
  doi       = {10.1016/j.autcon.2023.104885},
  publisher = {Elsevier BV},
}

@InProceedings{AkarshS.2022,
  author    = {M, Akarsh S. and Biradar, Rajashekar and Joshi, Prashanth V},
  booktitle = {2022 International Conference on Inventive Computation Technologies (ICICT)},
  title     = {Implementation of an Active Noise Cancellation Technique using Deep Learning},
  year      = {2022},
  month     = jul,
  publisher = {IEEE},
  doi       = {10.1109/icict54344.2022.9850807},
}

@Misc{Cosentino2020,
  author    = {Cosentino, Joris and Pariente, Manuel and Cornell, Samuele and Deleforge, Antoine and Vincent, Emmanuel},
  title     = {LibriMix: An Open-Source Dataset for Generalizable Speech Separation},
  year      = {2020},
  copyright = {arXiv.org perpetual, non-exclusive license},
  doi       = {10.48550/ARXIV.2005.11262},
  publisher = {arXiv},
}

@InProceedings{Dang2023,
  author     = {Dang, Dinh Son and Nguyen, Tung Lam and Ta, Bao Thang and Nguyen, Tien Thanh and Nguyen, Thi Ngoc Anh and Le, Dang Linh and Le, Nhat Minh and Do, Van Hai},
  booktitle  = {INTERSPEECH 2023},
  title      = {LightVoc: An Upsampling-Free GAN Vocoder Based On Conformer And Inverse Short-time Fourier Transform},
  year       = {2023},
  month      = aug,
  publisher  = {ISCA},
  series     = {interspeech 2023},
  collection = {interspeech 2023},
  doi        = {10.21437/interspeech.2023-677},
}

@Misc{Zhao2023,
  author    = {Zhao, Shengkui and Ma, Yukun and Ni, Chongjia and Zhang, Chong and Wang, Hao and Nguyen, Trung Hieu and Zhou, Kun and Yip, Jiaqi and Ng, Dianwen and Ma, Bin},
  title     = {MossFormer2: Combining Transformer and RNN-Free Recurrent Network for Enhanced Time-Domain Monaural Speech Separation},
  year      = {2023},
  copyright = {Creative Commons Attribution Non Commercial Share Alike 4.0 International},
  doi       = {10.48550/ARXIV.2312.11825},
  publisher = {arXiv},
}

@InProceedings{Liu2022,
  author     = {Liu, Haohe and Choi, Woosung and Liu, Xubo and Kong, Qiuqiang and Tian, Qiao and Wang, DeLiang},
  booktitle  = {Interspeech 2022},
  title      = {Neural Vocoder is All You Need for Speech Super-resolution},
  year       = {2022},
  month      = sep,
  publisher  = {ISCA},
  series     = {interspeech 2022},
  collection = {interspeech 2022},
  doi        = {10.21437/interspeech.2022-11017},
}

@Misc{Porr2020,
  author    = {Porr, Bernd and Daryanavard, Sama and Bohollo, Lucía Muñoz and Cowan, Henry and Porr, Bernd and Dahiya, Ravinder},
  title     = {Real-time noise cancellation with Deep Learning},
  year      = {2020},
  copyright = {arXiv.org perpetual, non-exclusive license},
  doi       = {10.48550/ARXIV.2011.03466},
  publisher = {arXiv},

}

@Misc{DellaLibera2022,
  author    = {Della Libera, Luca and Subakan, Cem and Ravanelli, Mirco and Cornell, Samuele and Lepoutre, Frédéric and Grondin, François},
  title     = {Resource-Efficient Separation Transformer},
  year      = {2022},
  copyright = {arXiv.org perpetual, non-exclusive license},
  doi       = {10.48550/ARXIV.2206.09507},
  publisher = {arXiv},
}

@Misc{Li2022,
  author    = {Li, Chenda and Yang, Lei and Wang, Weiqin and Qian, Yanmin},
  title     = {SkiM: Skipping Memory LSTM for Low-Latency Real-Time Continuous Speech Separation},
  year      = {2022},
  copyright = {arXiv.org perpetual, non-exclusive license},
  doi       = {10.48550/ARXIV.2201.10800},
  publisher = {arXiv},
}

@Misc{AlBadawy2021,
  author    = {AlBadawy, Ehab A. and Gibiansky, Andrew and He, Qing and Wu, Jilong and Chang, Ming-Ching and Lyu, Siwei},
  title     = {VocBench: A Neural Vocoder Benchmark for Speech Synthesis},
  year      = {2021},
  copyright = {Creative Commons Attribution Non Commercial Share Alike 4.0 International},
  doi       = {10.48550/ARXIV.2112.03099},
  publisher = {arXiv},
}

@Misc{Wichern2019,
  author    = {Wichern, Gordon and Antognini, Joe and Flynn, Michael and Zhu, Licheng Richard and McQuinn, Emmett and Crow, Dwight and Manilow, Ethan and Roux, Jonathan Le},
  title     = {WHAM!: Extending Speech Separation to Noisy Environments},
  year      = {2019},
  copyright = {arXiv.org perpetual, non-exclusive license},
  doi       = {10.48550/ARXIV.1907.01160},
  publisher = {arXiv},
}

@Misc{Maciejewski2019,
  author    = {Maciejewski, Matthew and Wichern, Gordon and McQuinn, Emmett and Roux, Jonathan Le},
  title     = {WHAMR!: Noisy and Reverberant Single-Channel Speech Separation},
  year      = {2019},
  copyright = {arXiv.org perpetual, non-exclusive license},
  doi       = {10.48550/ARXIV.1910.10279},
  publisher = {arXiv},
}

@Article{Michelsanti2021,
  author    = {Michelsanti, Daniel and Tan, Zheng-Hua and Zhang, Shi-Xiong and Xu, Yong and Yu, Meng and Yu, Dong and Jensen, Jesper},
  journal   = {IEEE/ACM Transactions on Audio, Speech, and Language Processing},
  title     = {An Overview of Deep-Learning-Based Audio-Visual Speech Enhancement and Separation},
  year      = {2021},
  issn      = {2329-9304},
  pages     = {1368--1396},
  volume    = {29},
  doi       = {10.1109/taslp.2021.3066303},
  publisher = {Institute of Electrical and Electronics Engineers (IEEE)},
}

@Article{Stoller2018,
  author    = {Stoller, Daniel and Ewert, Sebastian and Dixon, Simon},
  title     = {Wave-U-Net: A Multi-Scale Neural Network for End-to-End Audio Source Separation},
  year      = {2018},
  copyright = {arXiv.org perpetual, non-exclusive license},
  doi       = {10.48550/ARXIV.1806.03185},
  publisher = {arXiv},
}

@Article{Tzinis2020,
  author    = {Tzinis, Efthymios and Wang, Zhepei and Smaragdis, Paris},
  title     = {Sudo rm -rf: Efficient Networks for Universal Audio Source Separation},
  year      = {2020},
  copyright = {arXiv.org perpetual, non-exclusive license},
  doi       = {10.48550/ARXIV.2007.06833},
  publisher = {arXiv},
}

@InProceedings{Chhetri2023,
  author    = {Chhetri, Siddharth and Joshi, Manjusha Sanjeev and Mahamuni, Chaitanya Vijaykumar and Sangeetha, Repana Naga and Roy, Tushar},
  booktitle = {2023 2nd International Conference on Edge Computing and Applications (ICECAA)},
  title     = {Speech Enhancement: A Survey of Approaches and Applications},
  year      = {2023},
  month     = jul,
  publisher = {IEEE},
  doi       = {10.1109/icecaa58104.2023.10212180},
}

@Misc{Dubey2023,
  author    = {Dubey, Harishchandra and Aazami, Ashkan and Gopal, Vishak and Naderi, Babak and Braun, Sebastian and Cutler, Ross and Ju, Alex and Zohourian, Mehdi and Tang, Min and Gamper, Hannes and Golestaneh, Mehrsa and Aichner, Robert},
  title     = {ICASSP 2023 Deep Noise Suppression Challenge},
  year      = {2023},
  copyright = {Creative Commons Attribution 4.0 International},
  doi       = {10.48550/ARXIV.2303.11510},
  publisher = {arXiv},
}

@Misc{Takahashi2017,
  author    = {Takahashi, Naoya and Mitsufuji, Yuki},
  title     = {Multi-scale Multi-band DenseNets for Audio Source Separation},
  year      = {2017},
  copyright = {arXiv.org perpetual, non-exclusive license},
  doi       = {10.48550/ARXIV.1706.09588},
  publisher = {arXiv},
}

@Misc{Reddy2019a,
  author    = {Reddy, Chandan K. A. and Beyrami, Ebrahim and Pool, Jamie and Cutler, Ross and Srinivasan, Sriram and Gehrke, Johannes},
  title     = {A scalable noisy speech dataset and online subjective test framework},
  year      = {2019},
  copyright = {arXiv.org perpetual, non-exclusive license},
  doi       = {10.48550/ARXIV.1909.08050},
  publisher = {arXiv},
}

@Misc{Schulman2017,
  author    = {Schulman, John and Wolski, Filip and Dhariwal, Prafulla and Radford, Alec and Klimov, Oleg},
  title     = {Proximal Policy Optimization Algorithms},
  year      = {2017},
  copyright = {arXiv.org perpetual, non-exclusive license},
  doi       = {10.48550/ARXIV.1707.06347},
  publisher = {arXiv},
}

@Misc{Pascual2017,
  author    = {Pascual, Santiago and Bonafonte, Antonio and Serrà, Joan},
  title     = {SEGAN: Speech Enhancement Generative Adversarial Network},
  year      = {2017},
  copyright = {arXiv.org perpetual, non-exclusive license},
  doi       = {10.48550/ARXIV.1703.09452},
  publisher = {arXiv},
}

@Misc{Choi2019,
  author    = {Choi, Hyeong-Seok and Kim, Jang-Hyun and Huh, Jaesung and Kim, Adrian and Ha, Jung-Woo and Lee, Kyogu},
  title     = {Phase-aware Speech Enhancement with Deep Complex U-Net},
  year      = {2019},
  copyright = {arXiv.org perpetual, non-exclusive license},
  doi       = {10.48550/ARXIV.1903.03107},
  publisher = {arXiv},
}

@Misc{Hu2020,
  author    = {Hu, Yanxin and Liu, Yun and Lv, Shubo and Xing, Mengtao and Zhang, Shimin and Fu, Yihui and Wu, Jian and Zhang, Bihong and Xie, Lei},
  title     = {DCCRN: Deep Complex Convolution Recurrent Network for Phase-Aware Speech Enhancement},
  year      = {2020},
  copyright = {arXiv.org perpetual, non-exclusive license},
  doi       = {10.48550/ARXIV.2008.00264},
  publisher = {arXiv},
}

@Misc{Park2016,
  author    = {Park, Se Rim and Lee, Jinwon},
  title     = {A Fully Convolutional Neural Network for Speech Enhancement},
  year      = {2016},
  copyright = {arXiv.org perpetual, non-exclusive license},
  doi       = {10.48550/ARXIV.1609.07132},
  publisher = {arXiv},
}

@Misc{Hao2020,
  author    = {Hao, Xiang and Su, Xiangdong and Horaud, Radu and Li, Xiaofei},
  title     = {FullSubNet: A Full-Band and Sub-Band Fusion Model for Real-Time Single-Channel Speech Enhancement},
  year      = {2020},
  copyright = {arXiv.org perpetual, non-exclusive license},
  doi       = {10.48550/ARXIV.2010.15508},
  publisher = {arXiv},
}

@Misc{Hershey2015a,
  author    = {Hershey, John R. and Chen, Zhuo and Roux, Jonathan Le and Watanabe, Shinji},
  title     = {Deep clustering: Discriminative embeddings for segmentation and separation},
  year      = {2015},
  copyright = {arXiv.org perpetual, non-exclusive license},
  doi       = {10.48550/ARXIV.1508.04306},
  publisher = {arXiv},
}

@article{Wissam_IET_Signal_Process2022,
  author = {Jassim, Wissam A. and Skoglund, Jan and Chinen, Michael and Hines, Andrew},
  title = {Speech quality assessment with WARP-Q: From similarity to subsequence dynamic time warp cost},
  journal = {IET Signal Processing},
  volume = {n/a},
  number = {n/a},
  pages = {},
  doi = {https://doi.org/10.1049/sil2.12151},
  url = {https://ietresearch.onlinelibrary.wiley.com/doi/abs/10.1049/sil2.12151},
  eprint = {https://ietresearch.onlinelibrary.wiley.com/doi/pdf/10.1049/sil2.12151},
 }

@inproceedings{Pariente2020Asteroid,
    title={Asteroid: the {PyTorch}-based audio source separation toolkit for researchers},
    author={Manuel Pariente and Samuele Cornell and Joris Cosentino and Sunit Sivasankaran and
            Efthymios Tzinis and Jens Heitkaemper and Michel Olvera and Fabian-Robert Stöter and
            Mathieu Hu and Juan M. Martín-Doñas and David Ditter and Ariel Frank and Antoine Deleforge
            and Emmanuel Vincent},
    year={2020},
    booktitle={Proc. Interspeech},
}

@book{DSP_book,
author = {Lyons, Richard G.},
title = {Understanding Digital Signal Processing},
year = {1996},
isbn = {0201634678},
publisher = {Addison-Wesley Longman Publishing Co., Inc.},
address = {USA},
edition = {1st},
}

@ARTICLE{Younkin08,
  author={Younkin, Audrey C. and Corriveau, Philip J.},
  journal={IEEE Transactions on Broadcasting}, 
  title={Determining the Amount of Audio-Video Synchronization Errors Perceptible to the Average End-User}, 
  year={2008},
  volume={54},
  number={3},
  pages={623-627},
  doi={10.1109/TBC.2008.2002102}}
\end{document}